\documentclass[a4paper]{jpconf}
\usepackage{graphicx}

\newcommand{\Pt}{p_{T}}
\newcommand{\Et}{E_{T}}
\newcommand{\Ptg}{p_{T}^{\gamma}}
\newcommand{\lt}{\!<\!}
\newcommand{\gt}{\!>\!}
\newcommand{\gpj}{``$\gamma+jet$''~}
\newcommand{\newscale}{p_T^\gamma f(y^\star)}
\newcommand{\Fystar}{([1+exp(-2|y^\star|)]/2)^{1/2}}
\newcommand{\ystar}{0.5(\eta^\gamma-\eta^{jet})}

\begin{document}
\title{Measurement of triple differential photon plus jet cross section by D\O \vspace*{1.0cm}}

\author{Ashish Kumar ({\it for the D\O\ collaboration})}

\address{Department of Physics, The State University of New York at Buffalo, NY 14260, USA.}

\ead{ashishk@fnal.gov}

\begin{abstract}
We report on a new measurement of triple differential cross section for 
the process $p\bar{p}\rightarrow \gamma\, +\,jet\, + \,X$  in $p\bar{p}$ collisions at $\sqrt{s}=$1.96 TeV by the D\O\
Collaboration at Fermilab based on dataset corresponding to an integrated luminosity of 1.1 fb$^{-1}$.
\end{abstract}

\section{Introduction}
The photon plus jet (\gpj) events have the sheer advantage that the 
isolated photons are mostly ``direct'' photons, which emerge unaltered from 
the hard-scattering and provide clean probe of the parton level dynamics \cite{JFOwens}.
Together with the jet kinematics, these events shed more light on the 
underlying QCD dynamics. The parton level subprocesses include the
Compton-scattering   $gq\to q\gamma $ which dominates in a wide kinematic 
range and the annihilation subprocess $q\bar{q}\to g\gamma $. 
Studies of \gpj events, therefore, offer precision tests of 
perturbative QCD as well as information on the gluon density 
inside the colliding hadrons. 
Photons from hadron ($\pi^0,\eta$, {\it etc.}) decays and the
bremsstrahlung process 
provide the major background particularly at low transverse momentum ($\Pt$). 
However, their contribution is suppressed after application of tight 
photon isolation criteria. 

Using about 1.1 fb$^{-1}$ of Tevatron Run II data, D\O\ has measured 
the triple  differential cross section 
for $p\bar{p}\rightarrow \gamma\, +\,jet\, + \,X$ process
with a photon in the central pseudorapidity region,
$|\eta^{\gamma}|<1.0$, and a leading jet with $p_T > 15$ GeV in 
either the central ($|\eta^{jet}|<0.8$)
or the forward ($1.5<|\eta^{jet}|<2.5$) region. The photon $\Pt$ ($\Ptg$) varies
from 30 to 300 (200) GeV for the central (forward) jets. 
The analysis considers four kinematic regions differing in  $|\eta^{\gamma}|$ 
and $|\eta^{jet}|$: (1) both central and same side, (2) both  central and 
opposite side, (3) central photon, forward jet and same side, 
and (3) central photon, forward jet and opposite side. The kinematic region
in the $x-Q^2$ plane covered by the analysis (approximately 0.007$\leq x \leq$
0.7 and 0.9$\times$10$^3 \leq Q^2 \leq 0.4 - 0.8 \times 10^5$ GeV$^2$) 
significantly extends previous measurements.

\section{Event Selection and Background Suppression}
Photons are identified in the D\O\ detector as isolated energy deposits 
in the electromagnetic (EM) calorimeter consisting of 4 layers, EM1-EM4. 
Photon candidates are reconstructed with a cone algorithm with 
cone size 
${\mathcal
R}=\sqrt{(\Delta\eta)^2+(\Delta\phi)^2}=0.2$.  
Candidates
are selected if there is significant energy in the EM 
layers ($>96$\%), and the probability to have a matched track is less
than $0.1$\%, and they satisfy an isolation requirement
$[E_{total}(0.4)-E_{EM}(0.2)]/E_{EM}(0.2)<0.07$, where
$E_{total}(0.4)$ and $E_{EM}(0.2)$ are the total and EM energies 
within cone size of 0.4 and 0.2, respectively. We also limit the 
energy weighted cluster width in the finely-segmented EM3 layer.
Potential backgrounds from $W$ boson decays to electrons and cosmics
were suppressed by the cut on missing transverse 
energy $\Et^{miss}<12.5 + 0.36~\Ptg$. 
Three discriminating variables were used for further background suppression: the
number of EM1 cells with energy $>0.4$~GeV within
${\mathcal R}<0.2$, the  fraction of the EM cluster energy deposited 
in the EM1 layer , and the scalar sum of track $p_T$ in the 
ring of $0.05 \leq {\cal R} \leq 0.4$ (with $\Pt^{track}>0.4$ GeV) around 
the photon cluster direction.
These variables turned out to be very efficient for background suppression 
and show consistent behavior for MC/data electrons from $Z\to ee$ events. 
They are used as an input to an artificial neural network (NN) optimized 
for pattern recognition. An additional cut on the NN 
output $O_{\rm NN}>0.7$ is applied. The total photon selection 
efficiency is about 62\% at $\Ptg \simeq$
30 GeV and increases to a plateau of $\geq$70\% at $\Ptg \geq$70 GeV.
The event is required to have at least one hadronic jet reconstucted 
with a cone algorithm of ${\mathcal R}=0.7$.
The photon candidate and the most energetic jet were required 
to be well separated, $d{\mathcal R}(\gamma,jet)>0.7$. 
The total number of \gpj events remaining in Regions 1--4 after 
application of all the selection criteria is 
about 2.4 million ($\sim$34.4\%, 30.2\%, 20.1\% 13.3\% in Regions 1 to 4). 
These events are used to calculate the cross sections 
in 15 $\Ptg$ bins varied from 30 to 300 GeV
for Regions 1, 2 and in 13 $\Ptg$ bins varied from 30 to 200 GeV 
for Regions 3, 4. 

The selected signal sample still contains a sizeable background dominated 
by the di-jet events when one jet fluctuates to a
well isolated EM cluster. These jets are primarily composed of one or more
neutral mesons that decay into photons, and may also be accompanied by 
other soft hadrons whose energies
are deposited in the EM portion of the calorimeter.
Since the signal events cannot be identified on an event-by-event basis,
their fraction is determined statistically for a given $\Ptg$ bin.
The photon purity ($\mathcal P$) is determined by fitting the ANN 
distribution in data to a linear combination of the predicted ANN 
distributions for the signal and the background.

\section{Calculation of \gpj Cross Section and Comparison with Theory}
\begin{figure}[h]
\hspace*{-10mm} \includegraphics[width=14pc]{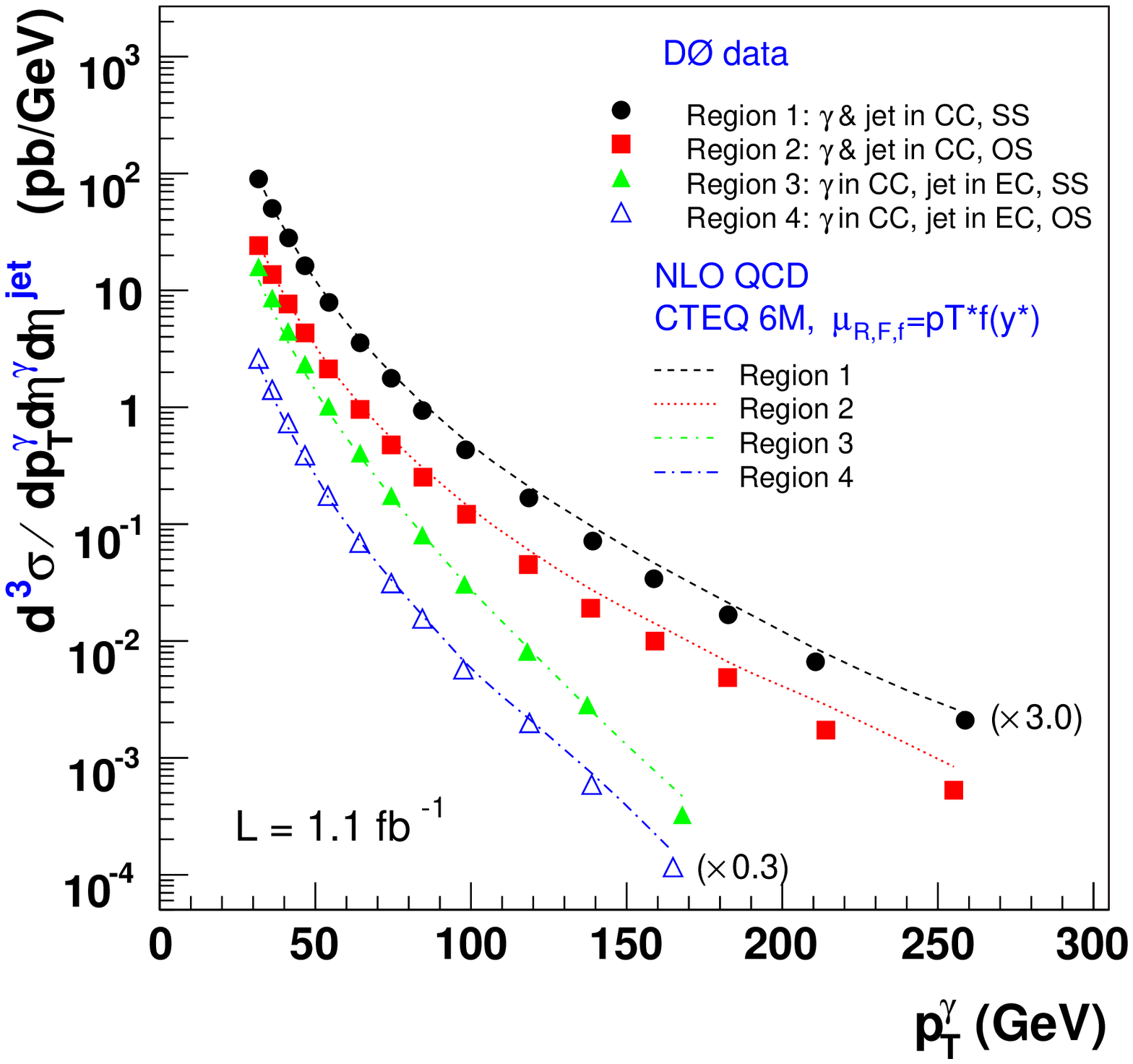}
\hspace*{-3mm} \includegraphics[width=14pc]{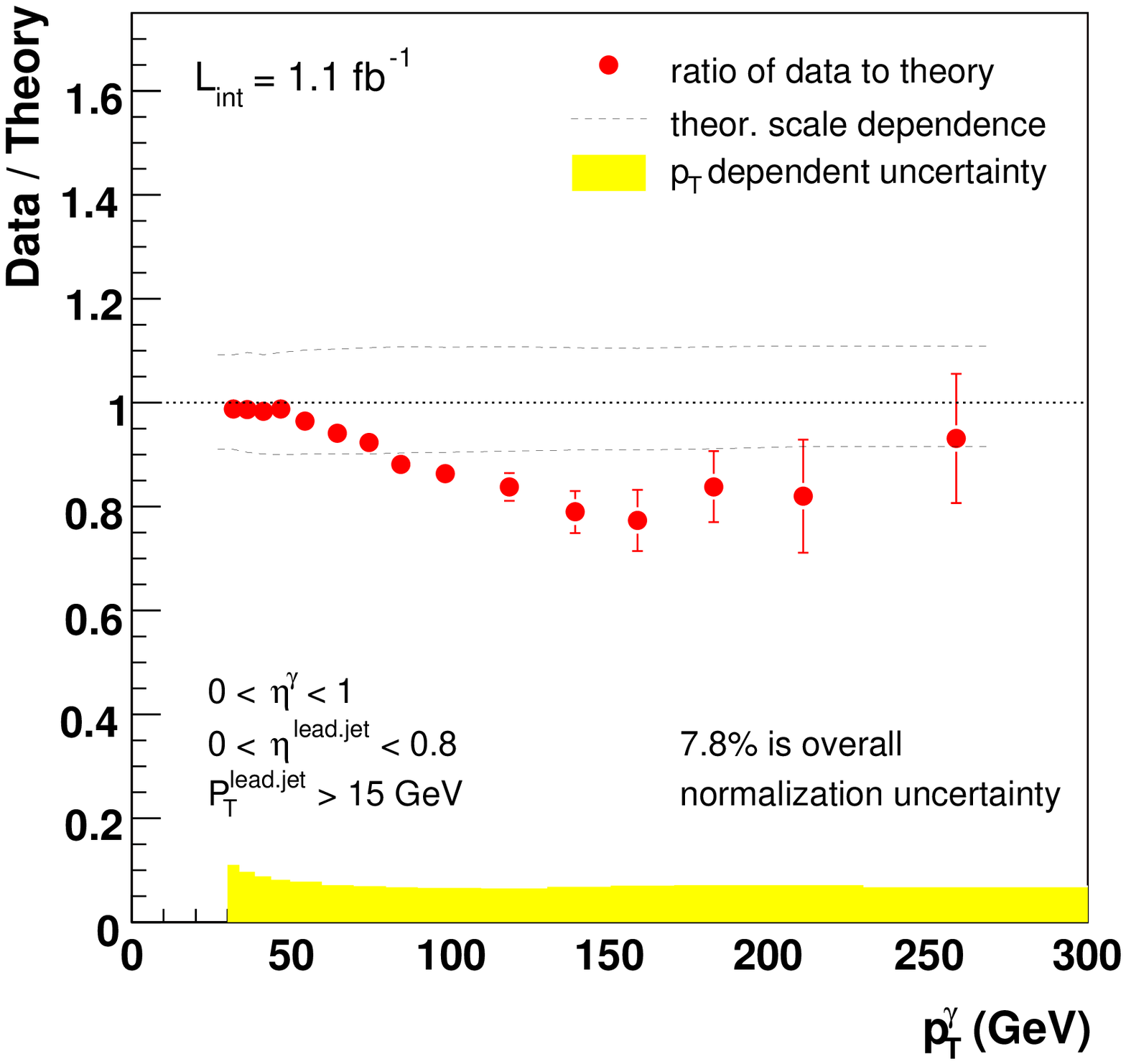}
\hspace*{-3mm} \includegraphics[width=14pc]{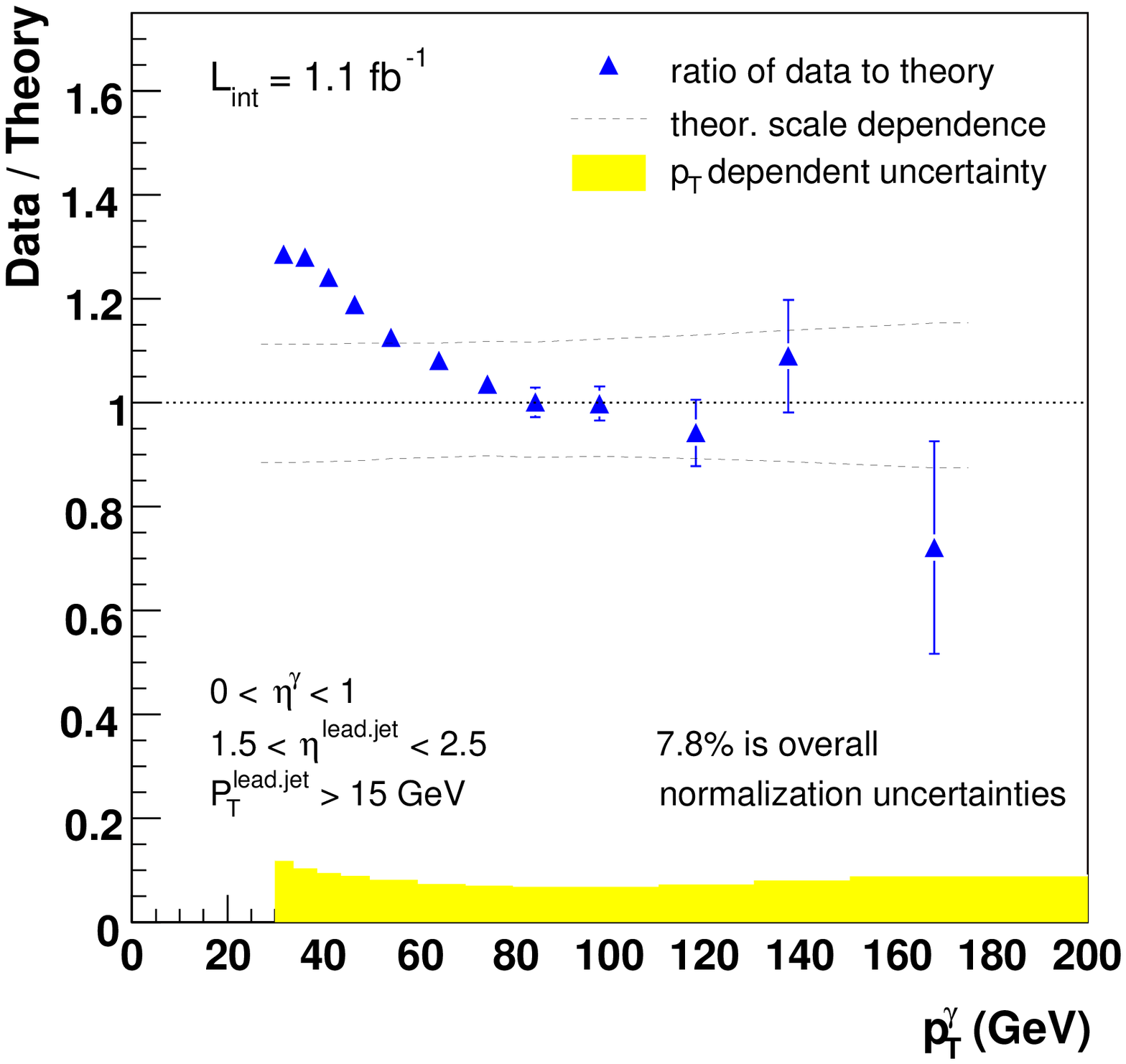}

\caption{{\sf Left}: The  $\Ptg$ spectrum of measured  \gpj cross sections for the four Regions along with the NLO predictions by {\sc jetphox} \cite{JETPHOX} shown as curves (they are scaled by factors 3.0 \& 0.3 for Regions 1 \& 4, respectively). 
The data are plotted at the $\Ptg$-weighted average of the fit function for each bin. {\sf Middle}: The ratio of the measured to the predicted cross section 
in {\it Region 1}. Dashed lines represent the effect of scale variations by factor of two. {\sf Right}: similar ratio but for {\it Region 3}. Refer to the text for the details.}
\label{fig:cross}
\end{figure}

The triple differential \gpj cross section is obtained using the relation:
\begin{eqnarray}
\frac{d^3\sigma}{d\Ptg ~d\eta^{\gamma}~d\eta^{jet}} = \frac{N ~{\cal P}~U}{L_{int} ~\Delta\Ptg ~\Delta\eta^{\gamma} ~\Delta\eta^{jet}
~\epsilon_{s}^{\gamma+jet}  }
\label{eq:cross}
\end{eqnarray}
\noindent
where 
$N$ is the number of \gpj candidates in the selected sample,
$L_{int}$ is the integrated luminosity,
$\epsilon_{s}^{\gamma+jet}$ is the signal selection efficiency, 
$\Delta \Ptg$, $\Delta \eta^{\gamma}$ and $\Delta \eta^{jet}$ are the 
bin sizes in $p_T^{\gamma}$, $\eta^{\gamma}$ 
and $\eta^{jet}$. The factor $U$
corrects the cross section for the finite resolution of
the calorimeter. The measured cross sections are shown 
in Fig.~\ref{fig:cross} as a function of $\Ptg$
with the full experimental (systematic $\oplus$ statistical) errors. 
The largest uncertainties are caused by the purity estimation, 
photon and jet selections and luminosity. Statistical errors vary 
from 0.1\% in the first $\Ptg$ bin to $13-20$\% in the last bin
while systematic errors are within $10-15$\% depending on region. 
The superimposed theoretical curve corresponds to the 
next-to-leading order (NLO) QCD predictions based 
on the {\sc jetphox} program \cite{JETPHOX} with the CTEQ6.1M 
set of parton distribution functions (PDF's) and BFG set of 
fragmentation functions \cite{CFGP_frag}. The choice of renormalization ($\mu_{R}$), 
factorization ($\mu_{F}$) and fragmentation ($\mu_{f}$) scales
used is $\mu_{R}=\mu_{F}=\mu_f=\newscale$ with $f({y^\star})=\Fystar$ 
and $y^\star=\ystar$.

The ratios of the measured cross sections to the NLO 
predictions are presented in Fig.~\ref{fig:cross} for the 
regions 1 and 3. The shape of data/theory in regions 2 and 4 is 
similar to that observed in regions 1 and 3, respectively. 
Only the statistical uncertainties are shown in the data points. 
The $p_T$ dependent systematic uncertainties are shown  on the plots 
separately by a shaded region. The scale dependence of the predictions is 
estimated by varying the scales by factors of two which yield 
uncertainties of 9--11\% for Regions 1--3 and 18--20\% for Region 4.  
The choice of different parameterizations of PDF's lead to 
variations in the data/theory ratios 
up to about 10\% for Regions 1--3 and up to $\sim$20\% for Region 4. 
The results of the measurements show a significant deviation 
from the NLO predictions for $\Ptg \gt 100$ GeV for the  
regions 1 and 2 and for $\Ptg \lt 50$ GeV for the Region 3. 
Such pattern of deviations in the data-to-theory ratios have earlier
been encountered by the UA2 \cite{UA2_phot}, CDF \cite{CDF_phot} 
and D\O\ \cite{Photon_paper_erratum} analyses. These results do
indicate the need of a better theoretical understanding 
of the processes with production of high energetic photons.
In particular, calculations enhanced for soft-gluon contributions
are expected to provide a better description of the data at low
$\Ptg$ in Region 3.

In order to reduce the systematic uncertainties,
we have also calculated ratios of the measured cross sections between different regions.  Most systematic uncertainties related to object ($\gamma$ \& jet) identification and luminosity are canceled in the ratios. The only systematic uncertainties that survive are related to the \gpj event purities
(since they differ a little between the four regions) and 
the jet selection efficiency when we calculate ratios with the central  jet in one region and the forward 
jet in another region. The overall experimental uncertainty estimated in such a way is about 3.5--9\% for
$44<\Ptg<110$ GeV and becomes larger for smaller $\Ptg$ (due to systematics) and larger  $\Ptg$ (due to statistics). Also, the scale uncertainty in the ratios is noticeably reduced; to the level of just 1--3\% for Regions 1--3 and 
up to 3--8\% for Region 4. In general, the shapes of the measured cross section ratios in data are qualitatively reproduced by the theory but 
we observe a quantitative disagreement for some kinematic regions even after taking into account the overall 
(experimental and theoretical scale) uncertainty. 
It is especially noticeable for the cross section ratios between Regions 1 \& 3 as well as between Regions 2 \& 3.

\end{document}